\documentclass[journal]{IEEEtran}

\usepackage{graphicx}
\usepackage{amssymb}
\usepackage{amsbsy}
\usepackage{ulem}
\usepackage{upgreek}
\usepackage{algorithm}
\usepackage{algorithmicx}
\usepackage{algpseudocode}
\usepackage{threeparttable}
\usepackage[cmex10]{amsmath}
\usepackage{array}
\usepackage[tight,footnotesize]{subfigure}
\usepackage{caption}
\usepackage{url}
\usepackage{epstopdf}
\usepackage{verbatim}
\usepackage{color}
\usepackage{float}
\usepackage{wrapfig}
\usepackage{xspace}

\usepackage{enumerate}

\newcommand*{\etc}{
    \@ifnextchar{.}
        {etc}
        {etc.\@\xspace}
}
\graphicspath{{figures/}}

\begin{document}

\newtheorem{thm}{Theorem}
\newtheorem{lma}{Lemma}
\newtheorem{defi}{Definition}
\newtheorem{proper}{Property}

\title{Maximizing Entanglement Routing Rate in Quantum Networks: Approximation Algorithms}

\author{Tu N. Nguyen,~\IEEEmembership{Senior Member,~IEEE}, Dung H. P. Nguyen, Dang H. Pham, Bing-Hong Liu, and Hoa N. Nguyen
\IEEEcompsocitemizethanks{
\IEEEcompsocthanksitem
T. N.~Nguyen is with the Department of Computer Science, Kennesaw State University, Marietta, GA 30060, USA (e-mail: tu.nguyen@kennesaw.edu).
\IEEEcompsocthanksitem
D. H. P. Nguyen and B.-H. Liu are with the Department of Electronic Engineering, National Kaohsiung University of Science and Technology, Kaohsiung 80778, Taiwan (e-mail: I109152111@nkust.edu.tw and bhliu@nkust.edu.tw).
\IEEEcompsocthanksitem
D. H. Pham and H. N. Nguyen are with the Department of Information Systems, VNU University of Engineering and Technology, Hanoi 10000, Vietnam (e-mail: dangph258@gmail.com and hoa.nguyen@vnu.edu.vn).
\IEEEcompsocthanksitem
Corresponding author: Tu N. Nguyen
}
}

\IEEEcompsoctitleabstractindextext{%
\begin{abstract}

There will be a fast-paced shift from conventional network systems to novel quantum networks that are supported by the quantum entanglement and teleportation, key technologies of the quantum era, to enable secured data transmissions in the next-generation of the Internet. Despite this prospect, migration to quantum networks cannot be done at once, especially on the aspect of quantum routing. In this paper, we study the maximizing entangled routing rate (MERR) problem. In particular, given a set of demands, we try to determine entangled routing paths for the maximum number of demands in the quantum network while meeting the network's fidelity. We first formulate the MERR problem using an integer linear programming (ILP) model to capture the traffic patent for all demands in the network. We then leverage the theory of relaxation of ILP to devise two efficient algorithms including HBRA and RRA with provable approximation ratios
for the objective function. To deal with the challenge of the combinatorial optimization problem in big scale networks, we also propose the path-length-based approach (PLBA) to solve the MERR problem.
Using both simulations and an open quantum network simulator platform to conduct experiments with real-world topologies and traffic matrices, we evaluate the performance of our algorithms and show up the success of maximizing entangled routing rate.
\end{abstract}
\begin{IEEEkeywords}
demand, entanglement, entangled pair, quantum networks, qubit.
\end{IEEEkeywords}}

\maketitle
\IEEEdisplaynotcompsoctitleabstractindextext
\IEEEpeerreviewmaketitle

\section{Introduction} \label{sec:intro}
The quantum mechanic appears as a promising technology to enable the secured quantum key distribution \cite{Bennett_2014}, distributed quantum computation \cite{10.1145/1412700.1412718,10.1145/3233188.3233224}, clock synchronization \cite{komar2014quantum}, and many other applications \cite{singh2021quantum}. The common characteristic of such applications is the requirement of \textit{entangled pairs} or \textit{EPR pairs} \cite{PhysRevLett.70.1895} between two distant quantum computers \cite{PhysRevLett.86.5188}. This is the reason that results in the emerging of quantum networks. The stages of the quantum network development were indicated together with several applications in \cite{doi:10.1126/science.aam9288}, and study \cite{10.1145/3341302.3342070} initially proposed a protocol for the link layer of the network. Furthermore, some networks have been constructed in several countries in practice \cite{10.1145/3387514.3405853,10.1145/3465481.3470056}, demonstrating the feasibility of the quantum network.

However, like other state-of-the-art technologies, the quantum network must cope with many challenges in the first growth stage \cite{8910635}. Two of them come from \textit{entanglement} and \textit{swapping} \cite{PhysRevLett.70.1895}, which are the fundamental steps to create an EPR pair between two \textit{end nodes} \cite{doi:10.1126/science.aam9288}, the stations where we run applications. Entanglement is the first step in the process of generating EPR pairs, where two \textit{quantum bits} \cite{10.1145/3341302.3342070} (or \textit{qubits} in short) in adjacent nodes establish entanglement states over \textit{quantum channels} to create \textit{quantum links}, also called the \textit{external links}. Each quantum link is generated successfully with a specific probability depending on many factors. One of which is the distance between two nodes; due to the inherent characteristic of signal amplitude decay along the transmission line, which is the optical fiber or free space in this situation, the longer distance, the lower the success probability. Fortunately, we can enhance the probability by attempting many photon pairs with different frequencies simultaneously; the external link will be generated when one out of the pairs establishes entanglement successfully \cite{9210823}. In addition, to overcome the difficulty of distance, one utilizes intermediate nodes, also called the \textit{quantum repeaters} \cite{10.1145/3341302.3342070,pirandola2017fundamental} (or \textit{repeaters} in short). In the second step, based on the end-to-end paths obtained from the routing, every two qubits inside the intermediate nodes swap together to create \textit{internal links}, which combines with the external links to form end-to-end entanglement. But as we will see in the next part, using too many repeaters will significantly affect the quality of entanglement. Both entanglement and swapping are probability processes. Therefore, the network is only determined after entanglement, and even when the network is identified, we still cannot guarantee the success of creating entangled pairs between two end nodes. In this paper, one node in the network can imply either an end node or a repeater. Due to the limitation of fabrication technology, quantum nodes are currently equipped with only a few qubits \cite{10.1145/3341302.3342070}, hence confining the capacity of the network. In addition, the network deployment challenges \cite{8910635} also affect the scalability of quantum network infrastructures such as quantum nodes and channels.

One way to increase the probability of success is to find multiple connections between two end nodes \cite{10.1145/3387514.3405853,li2021effective,pant2019routing}. Nevertheless, owing to the limitation of network resources, such as network infrastructure and the number of qubits inside a quantum node, this approach is not feasible in the case of multiple requests for generating entangled pairs at the same time. Because looking for multiple paths to meet one request may rapidly exhaust the network resources, which results in reducing network responsiveness. Instead, we are interested in leveraging the available resources to satisfy as many requests simultaneously as possible.

The \textit{fidelity} of an EPR pair is a vital metric that indicates the quality of entanglement. A higher fidelity, which means higher quality, results in a lower qubit error rate. The fidelity can alter from $0$ to $1$, but the desirable threshold is $0.5$. Error detection can work effectively with greater fidelity than this threshold \cite{10.1145/3341302.3342070,van2014quantum}. One factor influencing end-to-end fidelity is the length of the connection path between the two end nodes due to the degrading fidelity after each swapping \cite{9210823}. Therefore, to guarantee EPR pair quality to perform quantum applications, we need to confine the path length connecting two end nodes in terms of the number of hops. In fact, we can improve fidelity by entanglement purification \cite{doi:10.1126/science.aan0070}, but some types of quantum repeaters do not support this method. Moreover, this technique consumes network resources, where we have to sacrifice one or more quantum links.

On these bases, this study investigates the optimal techniques to meet as many demands of generating entangled pairs as possible while confining the length of the connecting path to guarantee the fidelity of the pairs.

\textbf{Motivation:}
Quantum networks face with the problem of forming entangled routing paths to connect demands (source-destination pairs) in the network.
A routing path in the conventional network can be constructed easily based on the topology of the network \cite{8556039,8422617}.
However, it is not the case in quantum networks since the routing paths must be entangled paths, so that we have to consider not only the conventional network resources (i.e., nodes and links) but also qubits and their entanglement relations \cite{10.1145/3387514.3405853,9210823}.
As the number of demands increases, the more number of entangled paths is needed as well. Hence, it is critical to perform \textit{dynamic traffic engineering}. To enhance the traffic engineering, it is important to enable maximum number of demands connected by entangled routing paths. In other worlds, we try to maximize the entangled routing rate for quantum networks.

\textbf{Challenges:} In summary, the maximizing entangled routing rate (MERR) problem has to address the following questions and aspects.

\begin{itemize}
    \item How to design a \textit{fast} algorithm to respond to the degradation of entangled links between qubits. In particular, the designed algorithm is expected to perform not only \textit{close optimal} but also \textit{fast} in terms of computation time.
    \item How many demands to connect in each period? Should their entangled routing paths are enabled at once?, and which specific qubits to be included in the paths? In addition, we have to ensure that each entangled routing path is short enough to meet the fidelity condition.

    \item Another challenge is that how to set up the real-world quantum network environment and then actually implement the proposed algorithm with appropriate metrics to evaluate its performance.

\end{itemize}

\textbf{Methodology and Contributions:}
We develop a methodology to address the above question posed by the next-generation network systems regarding entanglement routing. We introduce a model of qubit entanglement general enough to determine routing paths to connect demands. We then utilize this model to derive the optimal scheduling for qubit entanglement using the integer programming and different rounding techniques. The objective is mainly to maximize the entangled routing rate for the network. We show that the investigated problem is NP-hard and present the detailed algorithm with the best possible approximation ratio. We evaluate the performance of the proposed algorithms using both simulations and experiments that are implemented using the quantum network simulator, NetSquid \cite{netsquid}.

\begin{itemize}
    \item \textit{Maximizing Entangled Routing Rate Problem.} We introduce the problem of qubit entanglement scheduling, using general models of shared resources in quantum networks. Moreover, we also study the hardness of the MERR problem.
    \item \textit{Technique.} For the objective of
    maximizing entangled routing rate through the availability of qubit-enabled routing paths, we show that the optimization problem can be approached using the integer programming. We then present an approximation algorithm by using different rounding techniques.
    \item \textit{Real-world-based Evaluation.} We evaluate the proposed algorithms using real-world network topologies and traffic matrices.
    Evaluation results show that our approach can significantly increase the entangled routing rate for the network. Analyses of the proposed algorithms are also provided to verify their performance.

\end{itemize}

\textbf{Organization:} The rest of the paper is organized as follows. In $\S$\ref{sec:rlt_work}, we present the related studies on quantum network routing. In $\S$\ref{sec:model_problem}, we illustrates the system model and formally states the research problem. We describe our methods and provide the analysis of the algorithms in $\S$\ref{sec:method} and $\S$\ref{sec:analyze}, respectively. The experiment results are expressed in $\S$\ref{sec:sim}. We discuss the appropriation of the proposed techniques with a realistic network in $\S$\ref{sec:discussion}. Finally, in $\S$\ref{sec:conclusion}, we presents the conclusion and future works.

\section{Related Works}\label{sec:rlt_work}
Quantum applications have laid the foundation for the development of quantum networks. The emergence of the first quantum networks was intended for quantum key distribution (QKD) \cite{Bennett_2014}, a secure application in information exchange. These networks are presented in studies \cite{elliott2002building,Peev_2009,Sasaki:11}. This is inherently the beginning of the revolution. With the development of science and technology, quantum computers are gradually being realized \cite{8910635}, promising to exploit more quantum applications. This is also the basis for the strong development of quantum networks in the future.

Similar to the traditional network, the routing problem is one of the serious problems in the quantum network. Several works have initially studied this issue in different approaches. The first concern is the solution for increasing the success probability of entanglement and swapping, as well as ensuring the quality of entanglement. This problem can be addressed by finding the shortest path from the source node to the destination node of a demand. Research \cite{van2013path} is one of the first works to study the path selection in quantum networks using the traditional Dijkstra's algorithm. They define a link cost metric for the first time in quantum networks. They also present a deep analysis of the complexity of quantum networks compared with conventional networks. Study \cite{8068178} proposes a more complicated routing metric taking into account some physical factors, such as decoherence time and imperfect Bell state measurement. Both \cite{van2013path} and \cite{8068178} assume that there is only one request for generating the EPR pair at a specific time. This is basic but not practical because, along with the growth of the quantum network, the number of concurrent requests at a particular time will also increase significantly. Following the same approach of employing Dijkstra's algorithm, Pant \textit{et al.} \cite{pant2019routing} propose a greedy algorithm for multi-path routing, applying for grid networks after the entanglement. Distributing multi-path entanglement has also been investigated in \cite{li2021effective}, in which they propose three methods and evaluate the results according to such criteria as the traffic of the entire network, the degree of fairness between demands, and the traffic per edge in the network. They also carry out the routing after the entanglement and assess the results in a grid network. Vardoyan \textit{et al.} \cite{10.1145/3374888.3374899} study a star network to satisfy multiple users at the same time. They also propose a method to model the effect of decoherence in the network. Based on this model, the authors have evaluated the influence of the decoherence process on the performance of the system. In addition, Schoute \textit{et al.} \cite{schoute2016shortcuts} have investigated the routing problem of quantum networks but limited for ring and sphere topologies. Omar \textit{et al.} \cite{9259957} consider the routing in a grid network, taking into account trusted repeaters. Das \textit{et al.} \cite{PhysRevA.97.012335} assess entanglement routing in different particular topologies.
These above studies can be classified based on the following characteristics:

\begin{itemize}
    \item Some of them only looked at some networks with specific shapes. This is not consistent with the fact that the quantum network will have arbitrary shapes.
    \item The remaining works analyze many aspects of quantum networks without interesting in optimizing the network responsiveness.
\end{itemize}

In \cite{10.1145/3387514.3405853}, the authors suggest the technique of routing with two stages. The first stage happens before the entanglement, called the offline stage. In this stage, they predetermine some paths to serve the requests, including recovery paths for the case of failed entanglement. The second stage, named the online stage, happens after the entanglement. Based on the result of the entanglement and the predetermined paths in the offline stage, they determine the paths for the requests. In the situation that they cannot discover the paths for a request due to failed entanglement, they try to route through the recovery paths. Based on this direction, they presented an efficient method to distribute multiple entanglements simultaneously. Nevertheless, with this approach, they have to perform the routing twice; one is before the entanglement, and the other is after the entanglement. Moreover, this approach also cannot optimize the capacity of the network.

Several other studies look at the swapping process at quantum repeaters. By using the prepare and swap protocol, work \cite{9210823} propose a linear programming model applied to the multicommodity problem to optimize the entanglement distribution rate for the entire network with the requirements of ensuring the fidelity of demands. Study \cite{8967073} presents a more complicated protocol to optimize the entanglement distribution rate. However, this method requires a longer entangled state storage time.

Compared with the above studies, our work pays attention to practical and crucial problems in quantum networks, which is to optimize the ability to satisfy multiple demands at the same time. This issue stems from a limited character of the quantum network that is a quantum link that cannot be shared by many simultaneous demands. Our method can be applied to arbitrary networks. Indeed, the network that we use to perform the simulation is taken from traditional networks that have been deployed in practice.

\section{Network Model and Research Problem} \label{sec:model_problem}

\subsection{Quantum Network Model}\label{sec:basic}
We consider a centralized quantum network \cite{9351762}, in which a quantum node in the network consists of a traditional computer equipped with quantum processors and can interact with the other nodes via the quantum entanglement and teleportation.
Each node has a set of qubits (i.e., communication and data qubits).
The scope of this work focuses on discussing challenges of constructing entangled routing
paths for demands and how to overcome these challenges. Hereafter, let's imply that the following
discussion only refer to the communication qubits \cite{10.1145/3341302.3342070}.
Note that the number of qubits is limited in each node\footnote{It depends on the network infrastructure, then the number of qubits on each node is decided.}.
A node associated with qubits can execute entanglements with the other quantum nodes to form entangled links. In addition, one qubit can entangle to only one qubit at a time \cite{10.1145/3387514.3405853}. These two qubits are called Einstein–Podolsky–Rosen (EPR) pair which are in a maximally entangled state.

\begin{figure}[htbp]
	\centerline{\includegraphics[width=6cm]{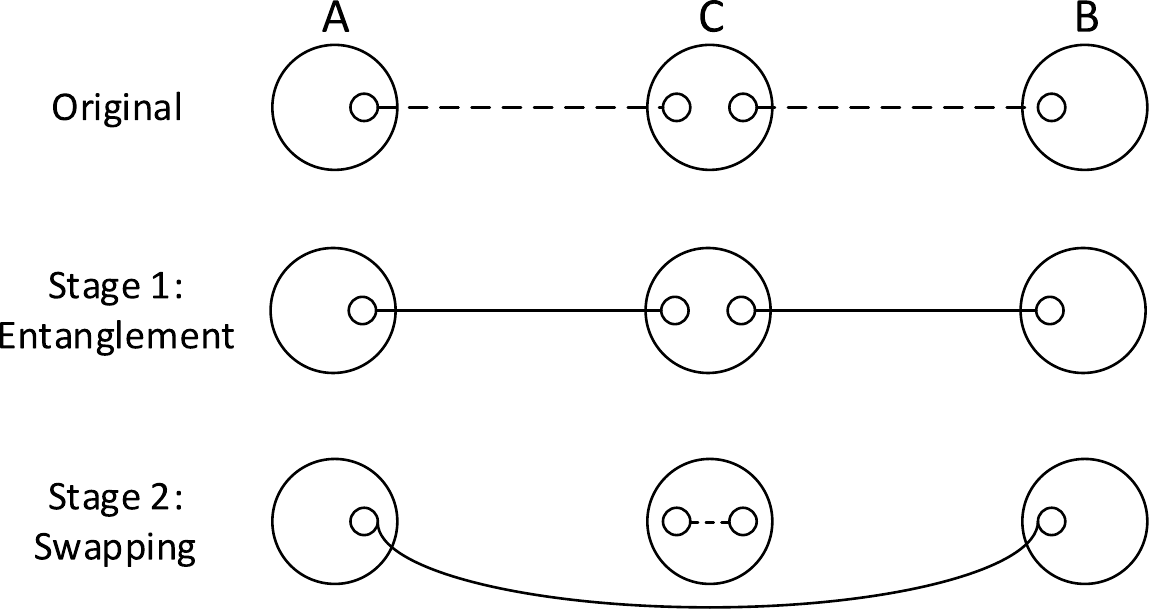}}
	\caption{Two-stage process.}
	\label{fig:two_pro}
\end{figure}

\begin{figure}[htbp]
\center
\subfigure[]{\includegraphics[width=4cm]{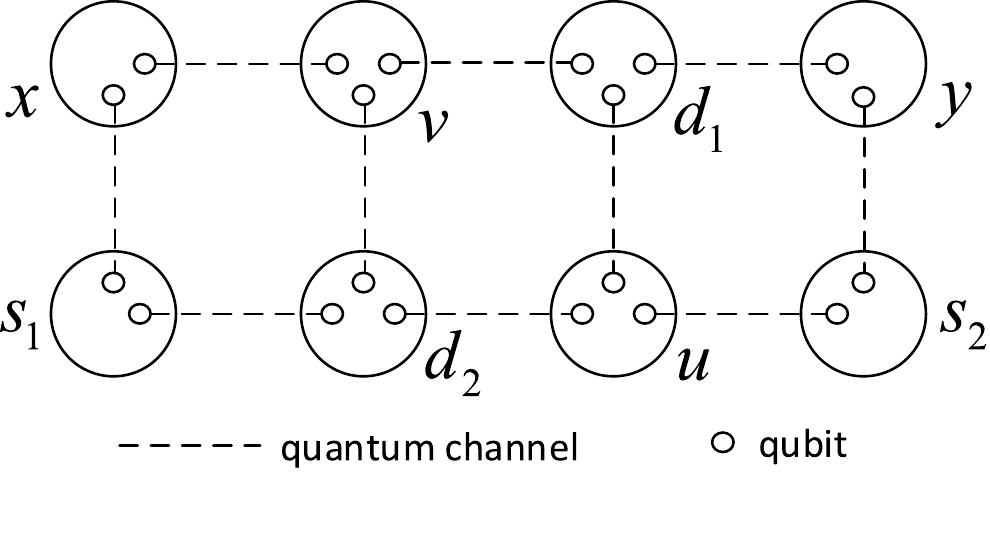}\label{fig:org_net}}
\subfigure[]{\includegraphics[width=4cm]{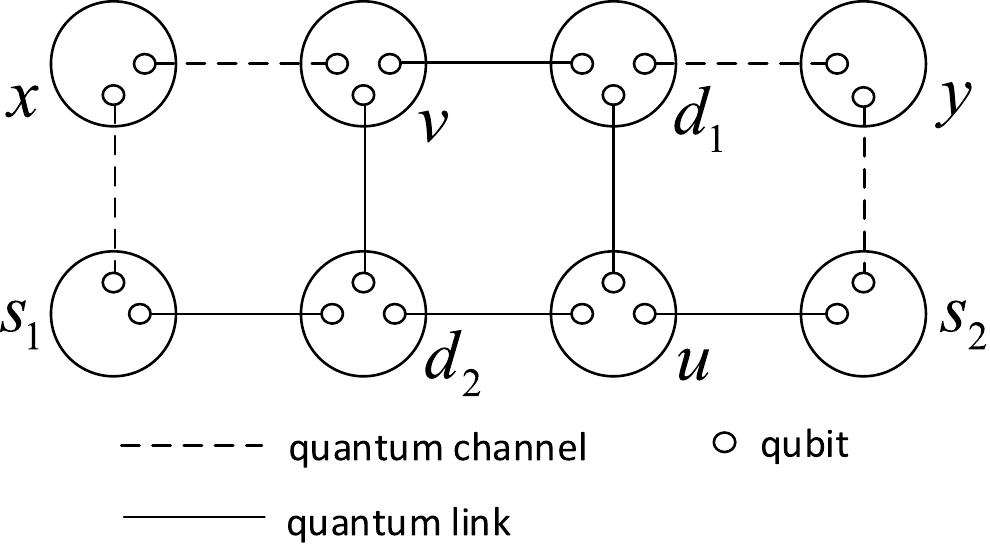}\label{fig:rsd_net}}
\caption{(a) Original network and (b) Reduced network}
\label{fig:network}
\end{figure}

In this paper, we employ a two-stage process to generate an EPR pair between two remote nodes, which are establishing entanglement between the qubits in neighbor nodes and swapping between the qubits at each node like Fig. \ref{fig:two_pro}. Both stages are successful with a specific probability depending on the physical layer. Because the entanglement may fail, the number of edges in the network after the entanglement can diminish compared with the original network, thus also called the \textit{reduced network}. Take the example from Fig. \ref{fig:network}, Fig. \ref{fig:org_net} displays the original network, which is the network before the entanglement. Assume edges $(s_1, x), (x, v), (d_1, y)$, and $(y, s_2)$ cannot establish entanglement, so we ignore these edges in the reduced network illustrated in Fig. \ref{fig:rsd_net}. Based on the reduced network, the controller will determine the route that connects two end nodes, and then repeaters in the network will perform the swapping depending on the controller decision. We also call a tuple consisting of two remote nodes, which require an EPR pair, a \textit{demand}.

Though the routing can be performed in the original network to mitigate the timing strictness, we have to reserve some paths for the case of failed entanglement like study \cite{10.1145/3387514.3405853}. This leads to two computing times; the first is for the routing in the original network, and the second is conducted after the entanglement to determine the appropriate recovery paths in the case of failure entanglement or even routing again in the situation that cannot find the recovery paths. This method is apparently not effective with respect to resource utilization and in the scenarios of the low success entanglement rate. That is the reason why we perform the routing in the reduced network.

\subsection{Research Problem}\label{sec:problem}
Given a quantum network described by an undirected graph $G = (V, E)$, in which $V$ and $E$ denote the sets of the vertices and edges, respectively. Each vertex in the graph is a quantum node (i.e., quantum computers or repeaters), and each edge is a quantum link.
The network traffic consists of a set of demands $D=\left\{ {{D}_{1}}, {{D}_{2}}, ..., {{D}_{n}} \right\}$ $\forall D_i = (s_i, d_i)$. An entangled routing path is defined as the routing path connecting the source and destination in which any link connects any two nodes is entangled link.
To form entangled routing paths connecting a source and a destination for a given demand, $(s_i, d_i)$, the network operator may decide to \textit{pair} qubits and makes these decisions among all demands.

\begin{defi}
Given a quantum network along with a set of demands (source-destination pairs).
The \textit{entangled routing rate} is defined as the number of demands that are connected via at least one entangled routing path over the total demands in the network.
\end{defi}

The entangled routing rate is apparently less than or equal to 1; the equality is equivalent to the best case once all input demands are met but rarely occurs owing to the reasons stated above. Thereby, we state the MERR problem as follows.

\textbf{INSTANCE:} Given a quantum network illustrated by an undirected graph $G = (V, E)$, where $V$ is the set of quantum nodes, $E$ is the set of quantum links, and a set of demands $D=\left\{ {{D}_{1}}, {{D}_{2}}, ..., {{D}_{n}} \right\}$, where $D_i = (s_i, d_i)$, and ${{l}_{\max }}\in {{\mathbb{Z}}^{+}}$, and a constant $k\in {\mathbb{Z}}^{+}$.

\textbf{QUESTION:} Does there exist a routing algorithm
to determine entangled routing paths connecting demands in the network such as the length of any path is not greater than $l_{max}$ and the total number demands connected by entangled routing paths is not less than $k$?

\begin{figure}[http]
\center
\subfigure[]{\includegraphics[width=4cm]{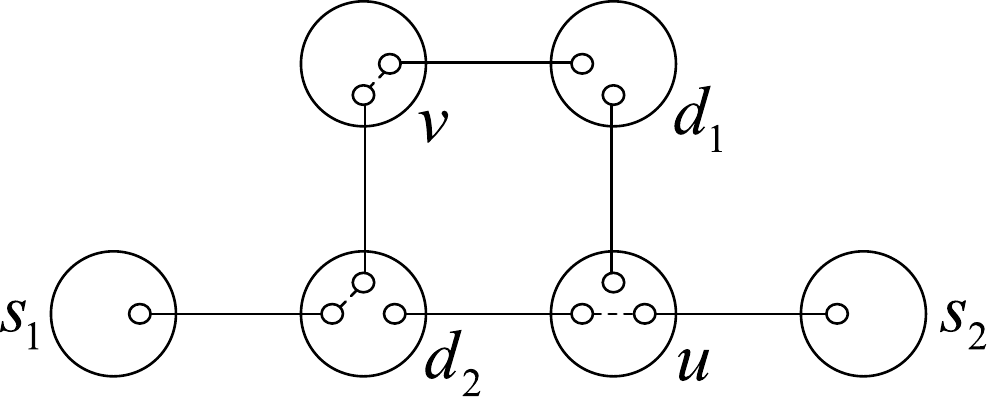}\label{fig:optimal_solution}}
\subfigure[]{\includegraphics[width=4cm]{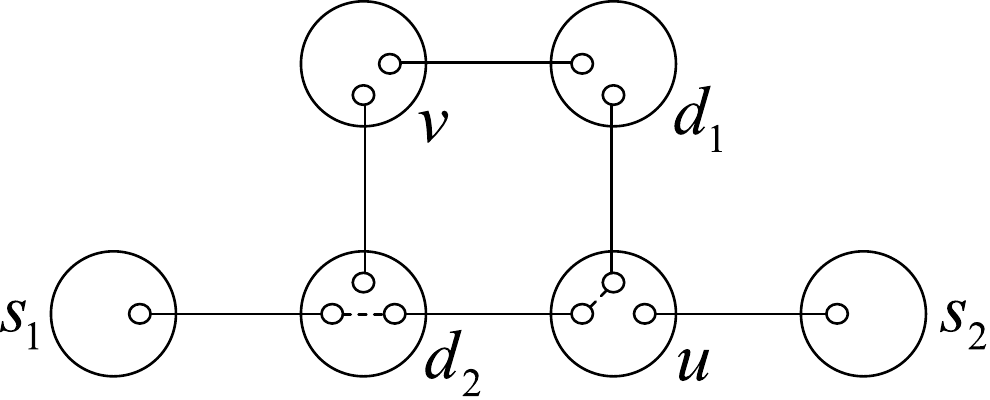}\label{fig:non_optimal_solution}}
\caption{Two methods to serve two demands ($s_1,d_1$) and ($s_2,d_2$): (a) Optimal solution and (b) Non optimal solution}
\label{fig:two_solution}
\end{figure}

Take the network in Fig. 2 as an example of the MERR problem, suppose we need to fulfill two demands ($s_1, d_1$) and ($s_2, d_2$) in the reduced network obtained from Fig. \ref{fig:rsd_net} simultaneously. Fig. \ref{fig:optimal_solution} shows the optimal solution by choosing path ($s_1, d_2, v, d_1$) for demand ($s_1, d_1$) and path ($s_2, u, d_2$) for demand ($s_2, d_2$), while Fig. \ref{fig:non_optimal_solution} displays the poor solution, where it chooses path ($s_1, d_2, u, d_1$) for demand ($s_1, d_1$), hence cannot meet demand ($s_2, d_2$) any more. The entangled routing rate is $1$ and $0.5$, respectively, in which the former is the optimal solution.

\subsection{Complexity Analysis of the MERR Problem}
In this section, we investigate thoroughly the \textit{complexity of the MERR problem}. In order to show the hardness of the MERR in Theorem \ref{thm:hardness}, our idea is to reduce the edge-disjoint path (EDP) problem to the MERR that is presented as follows.

\textbf{INSTANCE:} Given an undirected graph $\mathcal{G} = (\mathcal{V},\mathcal{E})$, where $\mathcal{V}$ and $\mathcal{E}$ are the set of vertices and edges, respectively, and a set of pairs of vertices $\left\{(\mathtt{s}_1,\mathtt{t}_1),\,(\mathtt{s}_2,\mathtt{t}_2),\,...,\, (\mathtt{s}_n,\mathtt{t}_n)\right\}$, where $\mathtt{s}_i,\,\mathtt{t}_i\in \mathcal{V},\,\forall{i}\,=\,1,\,2,\,...,\,n$, and a constant $\mathtt{h}\in {\mathbb{Z}}^{+}$.

\textbf{QUESTION:} Does there exist a method that connects each pair of the set with one path, such that the connected paths do not share their edges with the others and the number of connected paths is at least $\mathtt{h}$? 

\thm{The MERR problem is NP-hard.} \label{thm:hardness}

\textit{Proof:} Because the EDP problem is NP-hard \cite{1530717}, we prove the difficulty of the MERR problem by showing a reduction from the EDP problem to the MERR problem. Indeed, the instance of the MERR problem can be constructed from the instance of the EDP problem as follows: assign $V$ to $\mathcal{V}$, $E$ to $\mathcal{E}$, and set of demand $D$ to the set of pairs of vertices. This process is inherently done in polynomial time. To complete the proof, we demonstrate that the EDP problem is solvable if and only if the MERR problem is solvable.

Suppose that the EDP is solvable, which means there exists a constant $\mathtt{h}\in {\mathbb{Z}}^{+}$ such that the number of connected paths is at least $\mathtt{h}$. Therefore, without taking into account the constraint of path length, we will also determine at least $\mathtt{h}$ met demands for the MERR problem. Then, we drop out the demands whose connecting path is greater than $l_{max}$. Assume that the number of such demands is $\mathtt{h}'$. Hence, we can determine at least $\mathtt{h} - \mathtt{h}'$ met demands for the MERR problem.

Suppose that the MERR is solvable, we assign $l_{max}$ to a large enough value such as $\left|E\right|$. In this case, the solution to the MERR problem is also the solution to the EDP problem. This completes the proof.

\section{Tight Approximation for the MERR Problem} \label{sec:method}

In this section, we design tight approximation algorithms for the MERR problem by expressing the problem using the integer programming. We begin with the primary formulation of integer programming conditions in $\S$\ref{subsec:formulation} and detailed algorithms and routing techniques
are presented in the following.

\subsection{Primary Formulation}\label{subsec:formulation}
Theoretically, a quantum network is undirected, and thus, we could divide one undirected edge into two directed edges with opposite directions. On that basis, we convert the quantum network after the entanglement to a flow network, hence obeying the constraints of the flow network. Those are the flow conservation at the nodes and capacity limitation on the edges. Next, we consider each of the constraints in turn, in which symbol ${f_{uv}^{i}}$ indicates the flow of $i^{th}$ demand on directed edge $(u,v)$.

This paper only considers the single-channel networks, which implies each edge of the network possesses only one quantum channel. Because one quantum link cannot be shared between multiple paths of different demands at the same time, the edge’s capacity is only one. Hence, the first constraint is:
\begin{equation}\label{equ:cap_con}
    \begin{aligned}
        \sum\limits_{i=1}^{k}{f_{uv}^{i}}+\sum\limits_{i=1}^{k}{f_{vu}^{i}}\le 1
        & \,\,\,\,\,for\,each\,(u,v)\in E \\
    \end{aligned}
\end{equation}

At one node in the network, swapping between two internal qubits, with respect to the quantum, establishes entanglement between two qubits at two remote nodes that were being in the entanglement state with the internal qubits before. In terms of the flow network, the swapping at one node, but the source and destination, generates an internal link that connects the flow into the node and the flow out of that node. Hence, for each node in the network, but the source and destination, the total flow into the node must equal the total flow out of that node. Thus, we express the constraint following:
\begin{equation}\label{equ:conservation_con}
    \begin{aligned}
        \sum\limits_{\begin{smallmatrix}
        (u,v)\in E
        \end{smallmatrix}}{f_{uv}^{i}}-\sum\limits_{\begin{smallmatrix}
        (u,v)\in E
        \end{smallmatrix}}{f_{vu}^{i}}=0
        & \,\,\,for\,each\,u\in V-\{{{s}_{i}},{{d}_{i}}\} \\
        & \,\,\,and\,for\,each\,i=1,2,...,k \\
    \end{aligned}
\end{equation}

To avoid the situation where we can find a solution, but one out of the paths is too long, which results in unusable due to low fidelity and probability of success swapping, we pose the following constraint to limit the path length.
\begin{equation}\label{equ:length_con}
    \sum\limits_{(u,v)\in E}{\left( f_{uv}^{i}+f_{vu}^{i} \right)}\le {{l}_{\max }}\,\,\,for\,each\,i=1,2,...,k
\end{equation}
where $l_{\max}$ is the upper bound of the path length.

A demand is met when we can find a path to establish entanglement between two qubits in the source and destination, which means, concerning the flow network, there exists a flow from the source to the destination. Moreover, to save resources of the network to satisfy as many demands as possible, we consider only one path for each demand. Therefore, the total flow of $i^{th}$ demand out of $s_i$ is at most one, and the equality occurs only when the demand is met. This also implies the flows of $i^{th}$ demand into $s_i$ is equal to zero.
\begin{equation}\label{equ:src_flow_con}
    \sum\limits_{\begin{smallmatrix}
    ({{s}_{i}},v)\in E
    \end{smallmatrix}}{f_{{{s}_{i}}v}^{i}}\le 1\,\,\,for\,each\,i=1,...,k
\end{equation}
\begin{equation}\label{equ:src_inflow_con}
    \begin{aligned}
        f_{v{{s}_{i}}}^{i}=0
        & \,\,\,for\,each\,({{s}_{i}},v)\in E\,and \\
        & \,\,\,for\,each\,i=1,2,...,k \\
    \end{aligned}
\end{equation}

Finally, the flow of $i^{th}$ demand across directed edge $(u,v)$ in the network is a binary variable, in which one is equivalent to employing the corresponding edge to establish entanglement between the source and destination, and zero is vice versa.
\begin{equation}\label{equ:bin_con}
    \begin{aligned}
        f_{vu}^{i},\,f_{uv}^{i}\in \{0,1\}
        & \,\,\,\,\,for\,each\,(u,v)\in E\,and \\
        & \,\,\,\,\,for\,each\,i=1,2,...,k
    \end{aligned}
\end{equation}
\normalem

The goal of the MERR problem is to satisfy as many demands as possible at the same time. Because each met demand corresponds to a flow out of its source, the problem is equivalent to maximizing the total flows of all demands out of their corresponding source nodes. Therefore, the objective function of the problem is described as follows.
\begin{equation}\label{equ:obj1}
    \begin{aligned}
        & \textbf{P1}:\,maximize\,\,\,\,\,\,\sum\limits_{i=1}^{k}{\sum\limits_{\begin{smallmatrix}
        ({{s}_{i}},v)\in E
        \end{smallmatrix}}{f_{{{s}_{i}}v}^{i}}} \\
        & s.t.\,\,\,(\ref{equ:cap_con})-(\ref{equ:bin_con}) \\
    \end{aligned}
\end{equation}

In the following, we propose four approaches to address the MERR problem.
\subsection{Integer Linear Programming}
The first is to employ the integer linear programming (ILP) technique, in which the number of decision variables is $2\left| D \right|\left| E \right|$. Though this approach supplies the optimal solution, the running time will proliferate when the scale of the network increases. Therefore, this method is not applicable to large-scale networks because the quantum networks inherently require a short computation time to guarantee the coherence of qubits. Instead, it provides the upper bound reference with respect to the large networks. In the following, we propose three approximation algorithms to address the MERR problem in polynomial time.

\subsection{Half-Based Rounding}\label{subsec:hbr}
To reduce the complexity of the problem, we replace constrain (\ref{equ:bin_con}) with constraint (\ref{equ:relax}), thus changing the model from integer linear programming to linear programming, which is inherently solved within polynomial time. But most solutions obtained from this approach are not integers, thus not feasible because we cannot make decisions based on fractional values. Therefore, these values need to be rounded up to 1 or down to 0. Suppose the solution got from the linear program is $\overline{f_{uv}^{i}}$. If $\overline{f_{uv}^{i}}\ge 0.5$ then $f_{uv}^i = 1$, else $f_{uv}^i = 0$ if otherwise. Based on the rounding results, we choose the proper paths such that they satisfy all the constraints. Algorithm \ref{alg:hbr} describes our method.

\begin{equation}\label{equ:relax}
    \begin{aligned}
        f_{vu}^{i},\,f_{uv}^{i}\in \left[0,1\right]
        & \,\,\,\,\,for\,each\,(u,v)\in E\,and \\
        & \,\,\,\,\,for\,each\,i=1,2,...,k
    \end{aligned}
\end{equation}

In Algorithm \ref{alg:hbr}, the linear program is constructed and solved in Lines $1$–$2$. Then rounding stage and determining the satisfied demands are carried out over Lines $3$–$31$. After solving the linear program, some demands are satisfied immediately, which means some flow variables of these demands are equal to $1$, while the others are equal to $0$. Because these solutions met all constraints and are integers, we are certain to indicate the path from the source to the destination of these demands. They are then added into $S$, which is the set of met demands, and the remainder demands are put into set $R$. These steps are performed over Lines $3$-$10$. Afterward, in Lines $11$-$14$, the edges belonging to the paths of the met demands in set S are removed from the graph to guarantee the capacity constraint. Lines $15$-$23$ show the rounding stage for the remainder demands. The final solution ($f_{uv}^i$) will be rounded up to $1$, if the solution from the linear program ($\overline{f_{uv}^{i}}$) is greater than or equal to $0.5$, and down to $0$ if otherwise. The solutions from the rounding are employed to determine which demands are satisfied among the remaining demands in set $R$ (Lines $24$-$31$). The rounding strategy of this approach is based on the value of a half, thus called the Half-based Rounding Algorithm (HBRA).

\begin{algorithm}[http]
    \caption{Half-based Rounding Algorithm}\label{alg:hbr}
    \textbf{Input}: Network $G = (V,E)$, a set of demand $D = \{D_1, D_2, ..., D_n\}$ with $D_i = (s_i, d_i)$, ${{s}_{i}},{{d}_{i}}\in V$, a constraint of length $l_{\max}$ \\
    \textbf{Output}: A set of satisfied demands $S$
    \begin{algorithmic}[1]
        \State Build the linear program with objective function (\ref{equ:obj1}) and constraints (\ref{equ:cap_con}) - (\ref{equ:src_inflow_con}) and (\ref{equ:relax})
        \State Solve the program, the solution is expressed by $\overline{f_{uv}^{i}}$
        \State $S \gets \emptyset$, $R \gets \emptyset$
        \For {$D_i\in D$}
            \If {$D_i$ is satisfied} \Comment{Some demands can be satisfied after solving the linear program}
                \State $S\leftarrow S\cup {{D}_{i}}$
            \Else
                \State $R\leftarrow R\cup {{D}_{i}}$
            \EndIf
        \EndFor
        \For {$D_i\in S$}
            \State Determine the path $P_i$ from $s_i$ to $d_i$
            \State $E\leftarrow E\backslash \{e\in {{P}_{i}}\}$
        \EndFor
        \For {$D_i\in R$} \Comment{Rounding Stage}
            \For {$(u,v)\in E$}
                \If {$\overline{f_{uv}^{i}}\ge 0.5$}
                    \State $f_{uv}^{i}$ = 1
                \Else
                    \State $f_{uv}^{i}$ = 0
                \EndIf
            \EndFor
        \EndFor
        \For {$D_i\in R$}
            \If {Exist a path $P_i$ from $s_i$ to $d_i$}
                \If {$length({{P}_{i}})\le {{l}_{\max }}$}
                    \State $S\leftarrow S\cup {{D}_{i}}$
                    \State $E\leftarrow E\backslash \{e\in {{P}_{i}}\}$
                \EndIf
            \EndIf
        \EndFor
        \State \textbf{return} $S$
    \end{algorithmic}
\end{algorithm}

\subsection{Randomized Rounding}
This method is similar to HBRA, except for the rounding stage. Instead of using the threshold of a half, we utilize the technique proposed in \cite{raghavan1987randomized} to round the solution. Assume the solution obtained from the linear program is $\overline{f_{uv}^{i}}$, then the principle of the rounding is $Pr(f_{uv}^{i} = 1) = \overline{f_{uv}^{i}}$, where $Pr(f_{uv}^{i} = 1)$ is the probability of rounding $f_{uv}^{i}$ up to 1. The rounding is carried out only once to satisfy the strict timing of quantum networks. The paths are selected in the same way with the HBRA method. We call this approach the Randomized Rounding Algorithm (RRA).

\subsection{Path-Length-Based Algorithm (PLBA)}
In this approach, we apply the GREEDY\_PATH algorithm proposed by study \cite{10.1007/3-540-69346-7_12} to solve the MERR problem. This method is inspired by the idea that the shorter the path for one demand, the less impact on other demands. Namely, the algorithm tries to meet the demand owns the shortest path among the input demands first, hence called the Path-Length-Based Algorithm (PLBA). By removing the edges along the determined paths for the met demands, the algorithm guarantees to satisfy all the constraints. The core idea of this approach is the Dijkstra's algorithm, so it is a high-speed algorithm. The detail is presented in Algorithm \ref{alg:PLBA}, where we use the Dijkstra's algorithm to find the shortest path between two nodes $s_i$ and $d_i$ of demand $D_i$ in Line $11$. Then, we verify whether the algorithm can find the path or not, as well as the upper bound of the path length in Lines $12$-$15$. At each iteration in the while loop over Lines $6$-$17$, we fulfill the demand that has the shortest path. After that, in Lines $21$-$22$, we build the residual graph of $G$, in which the edges along the selected path are removed. We repeat this iteration until all demands are considered.

	\begin{algorithm}
		\caption{Path-Length-Based Algorithm}
		\textbf{Input}: Network Graph $G = (V, E)$, a set of demand $D = \{{D_{1}, D_{2}, ..., D_{n}}\}$ with $D_{i} = (s_{i}, d_{i})$, $s_i,\,d_i\in V$, a constraint of length $l_{max}$
		\label{alg:PLBA}
		
		\textbf{Output}: The number of met demands
		\begin{algorithmic}[1]
			\State $FUL\_D \gets \emptyset$ \Comment{List of fulfilled demands}
			\State $S \gets \emptyset$ \Comment{List of the shortest paths}
			\State $SPL \gets \emptyset$ \Comment{List of the shortest path length}
			\While{$len(FUL\_D) < len(D)$}
				\State $S \gets \emptyset$
			 	\While{$i < len(D)$}
			 		\If{$D_{i} \in FUL\_D$}
			 			\State $i \gets i+1$
			 			\State continue
			 		\EndIf
					\State $S_{i} \gets Shortest\_path(D_{i} = \{s_{i}, d_{i}\}) \:on \:G$
					\If{$len(S_{i}) > l_{max}$ or $len(S_{i}) = 0$}
						\State $SPL_{i} \gets 0$
						\State $FUL\_D \gets FUL\_D \cup D_{i}$
					\EndIf
					\State $i \gets i+1$
				\EndWhile
				\State Find $min(S_{j}) > 0$
				\State $SPL_{j} \gets len(S_{j})$
				\State $FUL\_D \gets FUL\_D \cup D_{j}$
				\State $R \gets G - \{Edges\,on\,the\,selected\,shortest\,path\,S_{i}\}$ \Comment{Build residual graph}
				\State $G \gets R$
			\EndWhile			
		\State \textbf{return} The number of $SPL_{j} > 0$
		\end{algorithmic}

	\end{algorithm}
\section{Analyze of the algorithms}\label{sec:analyze}
In this section, we first present the time complexity of the PLBA method over Theorem \ref{thm:plba_cplx}. In addition, Theorem \ref{thm:round_cplx} states the time complexity of the rounding stage of the HBRA and RRA algorithms. Then, Theorem \ref{thm:round_appr} demonstrates that the RRA technique can yield a feasible solution with a high probability.

\thm{The time complexity of the PLBA is $O({{\left| D \right|}^{2}}{{\left| V \right|}^{2}})$.}\label{thm:plba_cplx}

\textit{Proof:} The time complexity of the PLBA algorithm is $O(|D|^{2}|V|^{2})$, where $|D|$ is the number of demands, $|V|$ denotes the total number of nodes in the network. Indeed, in Line $11$, the PLBA uses Dijkstra's algorithm to find the shortest path between two nodes $(s_{i}, d_{i})$ of the demand $D_{i}$ with the worst-case complexity is $O(|V|^{2})$. In the \textit{while} loop from Line $6$ to Line $17$, we have to perform $|D|$ times Dijkstra's algorithm to find $|D|$ paths that satisfy $|D|$ demands, so we need $O(|D||V|^{2})$ operations. Next, the calculations through Lines $18$-$22$ take up negligible time. Finally, the while loop starting at Line 5 will run $|D|$ times. Consequently, the time complexity of the PLBA is $O(|D|^{2}|V|^{2})$.


\thm{The time complexity of the rounding stage in the HBRA and RRA algorithms is $O(\left|D\right|\left|E\right|)$.} \label{thm:round_cplx}

\textit{Proof:} Recall Algorithm \ref{alg:hbr}, the rounding stage is carried out through Lines 3 - 31. As interpreted in $\S$\ref{subsec:hbr}, $S$ is the set of demands that are satisfied immediately after solving the linear program. In the worst case, this set is empty because all solutions are fractional, which means that set $R$ is also set $D$. In this circumstance, the loop between Line 4 and Line 10 requires a time of at most $O(\left|D\right|)$. Because set $S$ is empty, the loop between Line 11 and Line 14 is ignored. Because set $R$ is also set $D$ in this case, the loop between Line 15 and Line 23 and the loop between Line 24 and Line 31 require at most $O(\left|D\right|\left|E\right|)$ and $O(\left|D\right|)$ time, respectively. Therefore, the time complexity of the rounding stage computing over the above loops is $O(\left|D\right|\left|E\right|)$.

\thm{Suppose that the optimal solution got from the linear programming relaxation is $\mathcal{O}_{LP}$, and the RRA technique yields a feasible solution $\mathcal{O}_{RRA}$, we have $\Pr \left[ {\mathcal{O}_{RRA}}\le (1-\varepsilon ){\mathcal{O}_{LP}} \right]\le {{e}^{-\frac{{{\varepsilon }^{2}}{\mathcal{O}_{LP}}}{2}}}$ with $0\le \varepsilon \le 1$.} \label{thm:round_appr}

\textit{Proof:} Assume that a solution got from the linear program is $\overline{f_{{{s}_{i}}v}^{i}}$, and the corresponding solution after the randomized rounding is $f_{{{s}_{i}}v}^{i}$. According to the definition of expectation, we have
\begin{gather*}
    E\left[ f_{{{s}_{i}}v}^{i} \right]=1\times \overline{f_{{{s}_{i}}v}^{i}}+0\times (1-\overline{f_{{{s}_{i}}v}^{i}})=\overline{f_{{{s}_{i}}v}^{i}}.
\end{gather*}
Therefore, by means of using the linearity of expectation, the expected value of $\mathcal{O}_{RRA}$ is
\begin{gather*}
    \begin{aligned}
    E\left[ {\mathcal{O}_{RRA}} \right]
    & =E\left[ \sum\limits_{i=1}^{k}{\sum\limits_{({{s}_{i}},v)\in E}{f_{{{s}_{i}}v}^{i}}} \right] \\
    & =\sum\limits_{i=1}^{k}{\sum\limits_{({{s}_{i}},v)\in E}{E[f_{{{s}_{i}}v}^{i}]}}=\sum\limits_{i=1}^{k}{\sum\limits_{({{s}_{i}},v)\in E}{\overline{f_{{{s}_{i}}v}^{i}}}}={\mathcal{O}_{LP}}.
    \end{aligned}
\end{gather*}
By applying Chernoff Bound \cite{dubhashi2009concentration} for $\mathcal{O}_{RRA}$, we have
\begin{gather*}
    \Pr \left[ {\mathcal{O}_{RRA}}\le (1-\varepsilon ){E\left[ {\mathcal{O}_{RRA}} \right]} \right]\le {{e}^{-\frac{{{\varepsilon }^{2}}{E\left[ {\mathcal{O}_{RRA}} \right]}}{2}}},
\end{gather*}
where $0\le \varepsilon \le 1$. Finally, by replacing $E\left[ \mathcal{O}_{RRA} \right]$ by $\mathcal{O}_{LP}$, we have
\begin{gather*}
    \Pr \left[ {\mathcal{O}_{RRA}}\le (1-\varepsilon ){\mathcal{O}_{LP}} \right]\le {{e}^{-\frac{{{\varepsilon }^{2}}{\mathcal{O}_{LP}}}{2}}}
\end{gather*}

\section{Experimental Evaluation}\label{sec:sim}

In this section, we first carry out experiments on quantum networks by employing an open platform to gain insight into the factors that impact the operation of the network. Then, we conduct simulations to evaluate the performance of the proposed methods as well as the influence of the infrastructure on the capacity of quantum networks.

\subsection{Simulation Settings}
In this paper, we employ two discriminate networks in reality collected by the Internet topology zoo \cite{6027859} to assess the performance of the proposed algorithms. The first one is the Surfnet topology with $50$ nodes and $68$ edges. The second is the US Carrier topology, in which the number of nodes and edges are $158$ and $189$, respectively. We also suppose the probability of entanglement is nearly $1$, which means most quantum links can be generated successfully. This assumption is reasonable because study \cite{9210823} showed the technique that boosts the success probability of entanglement. Hence, the amount of failed entanglement is negligible, and the network after entanglement is nearly similar to the original network. Therefore, these parameters also express the scale and resources of the networks, which are the significant factors associated directly with the number of the met demands. Next, all input demands are generated randomly based on these topologies such that the length of the shortest path from the source to the destination of each demand is less than or equal to $8$ hops. In the simulation, we let the number of input demands at the same time vary from $2$ to $20$. Moreover, to solve the linear programs in ILP, HBRA, and RRA approaches, we use the available LP solver in Python $3.9$ pulp class.

\begin{figure}[htbp]
\center
\subfigure[]{\includegraphics[width=4cm]{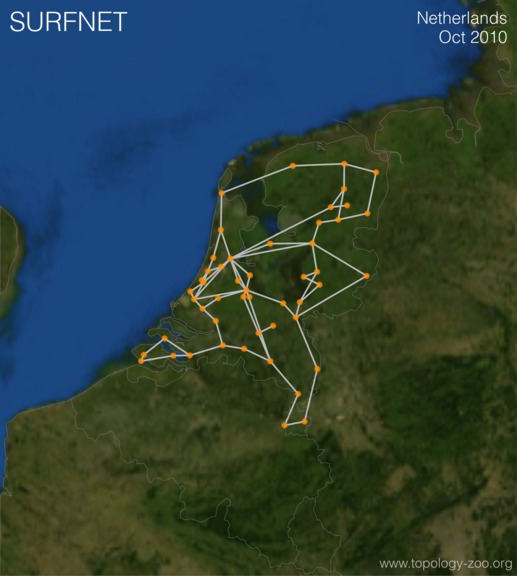}\label{fig:surf_topo}}
\subfigure[]{\includegraphics[width=4cm]{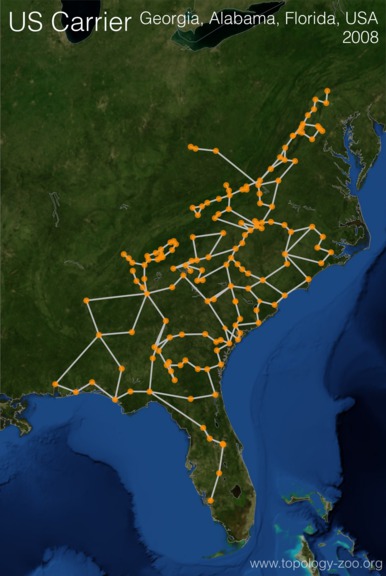}\label{fig:us_topo}}
\caption{Topologies for the simulations (a) Surfnet (b) US Carrier \cite{6027859}.}
\label{fig:topo}
\end{figure}

\subsection{Quantum Network Simulation}
To ensure the contribution of this work to the field, we carry out experiments using the open quantum simulator platform, NetSquid, to simulate the network environment using real-world topologies and traffic matrices to evaluate the performance of the proposed algorithms to solve the MERR problem. Thereby, we evaluate the appropriation of the proposed algorithms with the actual operation of the quantum network. First, we analyze the basic parameters of a quantum network. Two main characteristics of quantum network components are propagation delay and quantum error (quantum loss and quantum noise). Regarding about propagation delay, before generating entanglement between two nodes, in the middle of every quantum channel, there is a quantum source which emits EPR pairs by using Nitrogen-Vacancy (NV) centres in diamond \cite{childress_hanson_2013} to the both two nodes which have propagation delay $p_{d} = l_{d}/c_{f}$, where $l_{d}$(km) is the distance between two nodes; $c_{f}$ is the speed of light in a fiber optic cable and $c_{f}$ is calculated as $c_{v}/n_{r_i}$, where $c_{v}$ denotes the speed of light in vacuum ($30000$km/s) and the refractive index of the glass is $n_{ri} = 1.44$ i.e. $c_{f} = 300000/1.44 \approx 200000$(km/s). Next, two significant factors that result in quantum error are quantum loss and quantum noise. The difference between the two factors is that quantum loss is implemented before transmitting the message, while quantum noise is used just before receiving the message. The quantum loss is calculated by $p_{loss} = 1 - (1-p_{init}) * 10^{-\alpha d / 10}$, where $p_{init}$ is the probability that the qubit is lost immediately after generation process, $\alpha$ is the attenuation of an optical fiber (in dB/km) and $d$ is the distance which a photon travels. In quantum noise, a qubit is subject to decoherence through the dephasing noise and depolarizing noise. When applying the dephasing noise/depolarizing noise model, the noise will occur with probability $p$, and the qubit remains unchanged with probability $1-p$.
The dephasing noise is calculated with the probability $p_{deph} = 1 - \exp(-\Delta t R_{deph})$, where $R_{deph}$ is the dephasing rate (in Hz) and $\Delta t $ is the time delay (in ns). The probability of depolarizing is calculated as $p_{depo} = 1 - \exp(-\Delta t R_{depo})$, where $R_{depo}$ is the depolarizing rate (in Hz) and $\Delta t $ is the time delay (in ns).

\begin{figure}[htbp]
\center
\subfigure[]{\includegraphics[width=4cm]{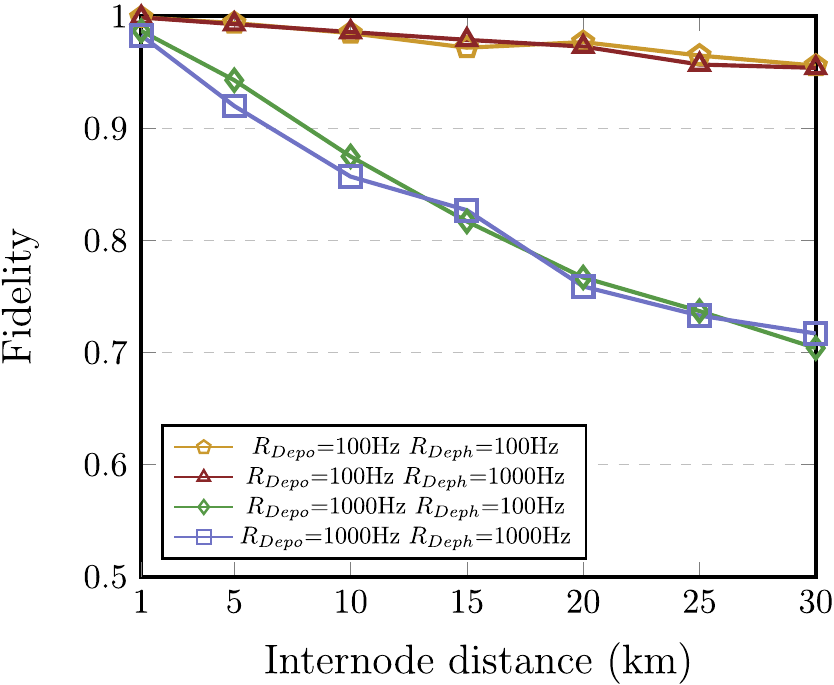}\label{fig:ev1}}
\subfigure[]{\includegraphics[width=4cm]{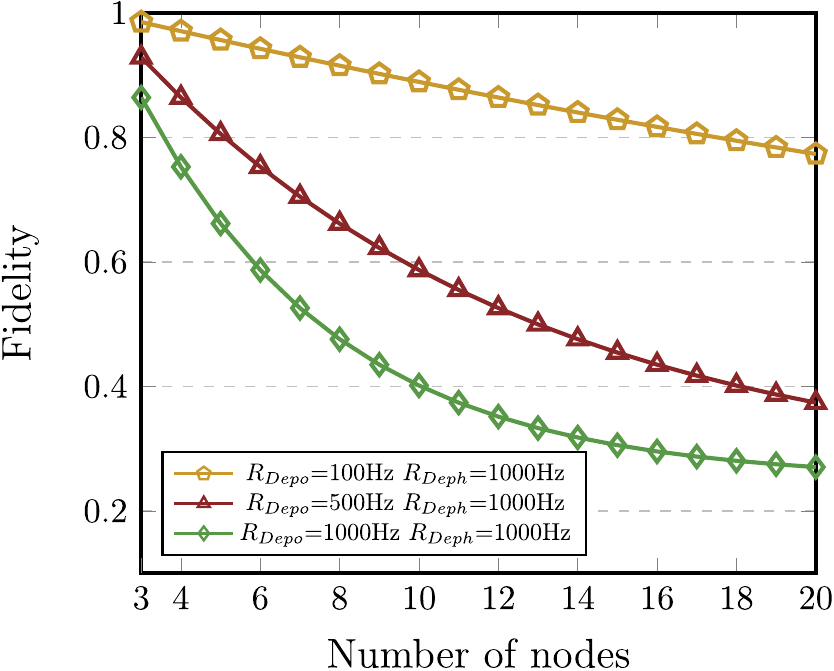}\label{fig:ev2}}
\caption{Impact of quantum noise on fidelity (a) Internode distance and (b) Number of nodes}
\label{fig:ev12}
\end{figure}

The first experiment we conduct is to consider the impact of quantum noise (dephasing rate and depolarizing rate) on the quality of teleportation when expanding the distance of internode (increased from $1$km to $30$km). We set up a quantum network with $p_{init} = 0$; the attenuation $\alpha = 0$dB/km; the dephasing rate $R_{deph} \in \{100,1000\}$Hz; the depolarizing rate $R_{depo} \in \{100,1000\}$Hz. As the result that is shown in Fig. \ref{fig:ev1}, it is obvious that the fidelity decreases gradually as the distance between two sender and receiver nodes increase. It can also be noted that there is a significant decline in the value of fidelity when the depolarizing rate rises, whereas the dephasing rate does not affect the fidelity much when we set a fixed depolarizing rate.

In the second experiment, therefore, we fix the value of $R_{deph}$ is $1000$Hz; $p_{init} = 0$; the attenuation $\alpha = 0$dB/km; various value of $R_{depo} \in \{100,5000,1000\}$Hz; the distance is kept at $20$km; to evaluate the impact of depolarizing rate on fidelity when we increase the number of nodes (from $3$ to $20$) in the repeater chain. From the results is illustrated in Fig. \ref{fig:ev2}, the fidelity drop remarkably depends on the increase of the depolarizing rate and the number of nodes.

\begin{figure}[htbp]
\center
\subfigure[]{\includegraphics[width=4cm]{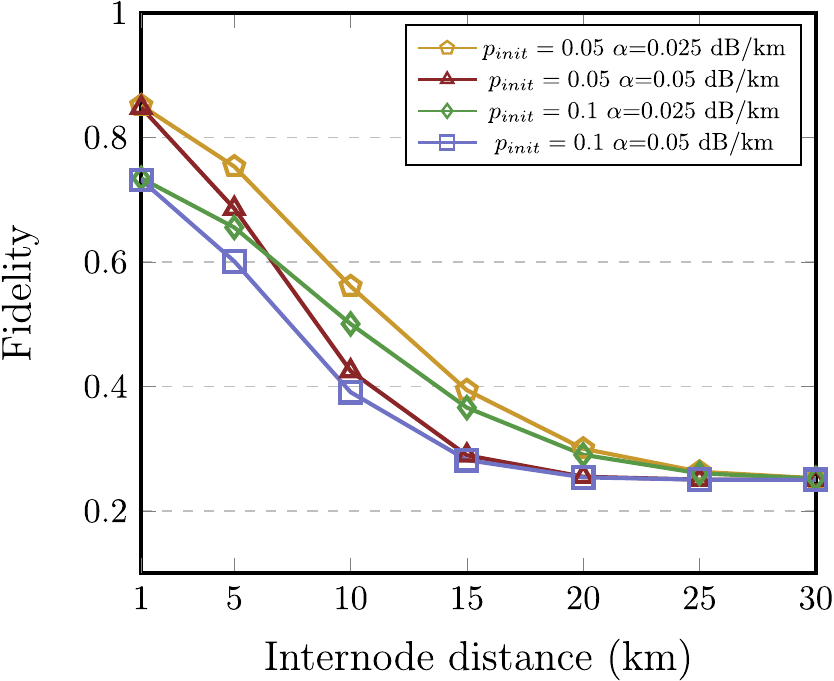}\label{fig:ev3}}
\subfigure[]{\includegraphics[width=4cm]{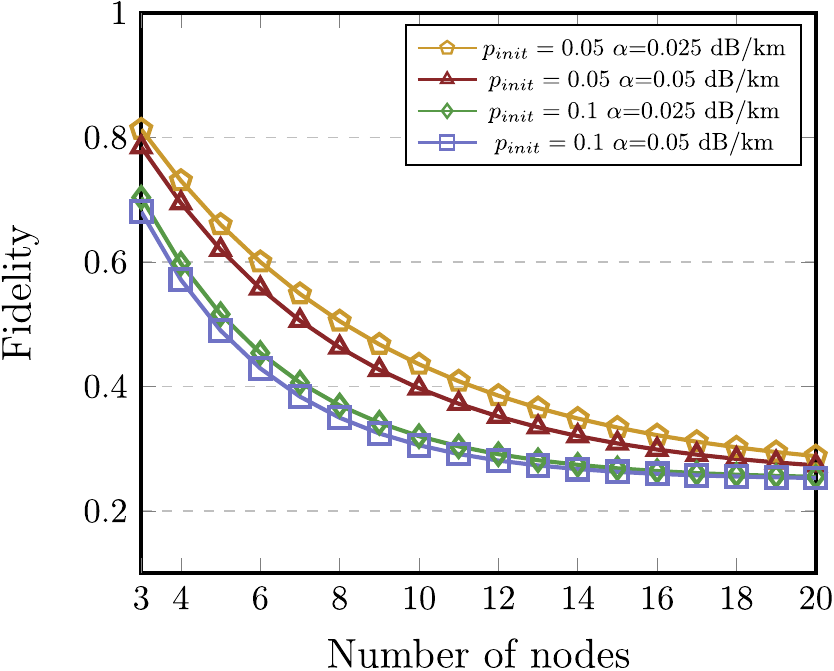}\label{fig:ev4}}
\caption{Effect of quantum loss on fidelity (a) Internode distance and (b) Number of nodes}
\label{fig:ev34}
\end{figure}

In the third and fourth experiments, we evaluate the effect of quantum loss on the quality of teleportation. We keep the constant value of $R_{depo}=1000$Hz, $R_{deph}=1000$Hz; assigned $p_{init} \in \{0.05, 0.1\}$; the attenuation $\alpha \in \{0.025, 0.05\}$dB/km. As we can see, the fidelity in both Fig. \ref{fig:ev3} and Fig. \ref{fig:ev4} tends to move down when increasing the distance of two nodes (from $1$km to $30$km) as well as the number of nodes in the quantum network (from $3$ to $20$).

\begin{figure}[htbp]
    \centering
    \includegraphics[width=8cm]{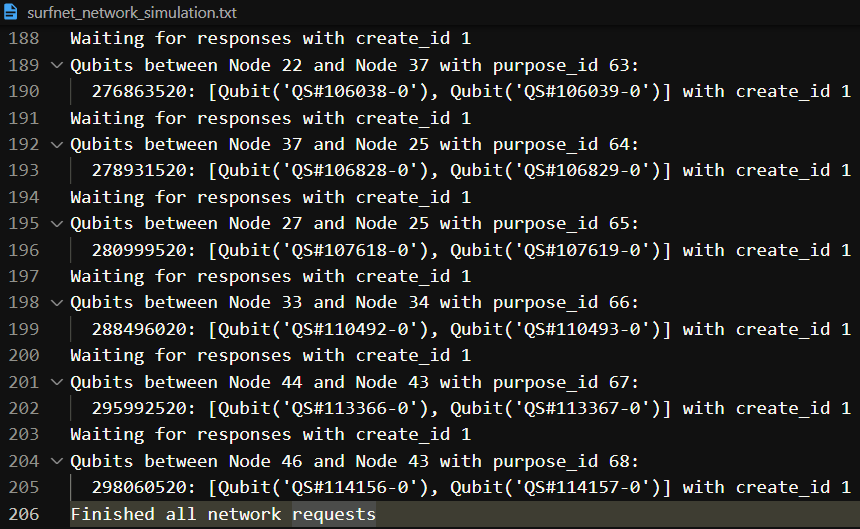}
    \caption{Required time to generate all entanglement in the Surfnet network.}
    \label{fig:ev56}
\end{figure}

In the fifth experiment, we simulate the entanglement process in the quantum network to evaluate the required time. The end of the log file for this simulation is depicted in Fig \ref{fig:ev56}. It takes a time of about $0.3$ seconds to create $68$ entangled pairs among $50$ nodes in the Surfnet network. Regarding the US Carrier network, it takes $0.77$ seconds to generate all entangled pairs. We will discuss more the required time in $\S$\ref{sec:discussion}.

\subsection{Performance Evaluation}
In the above sections, we propose four techniques to address the problem of optimizing the number of met demands. The first method is to employ an ILP model. This approach provides the optimal solution, but it is not appropriate with large-scale quantum networks due to the exponential computing time. Hence, we implement three approximation algorithms that can solve the problem in polynomial time, which can satisfy the strict timing of the quantum network. Two out of them utilize the technique of relaxing the integer linear program, then apply the rounding stage to obtain the feasible solution. The difference between them is the way of rounding, in which one employs a specific strategy (HBRA), and the other utilizes the randomized rounding method (RRA) proposed by \cite{raghavan1987randomized}. Finally, the remainder approximation method is a greedy algorithm based on the shortest path from the source to the destination of demands (PLBA). In this section, we evaluate the performance of the approximation approaches compared with the optimal method to gain insight into the proposed algorithms when applied to quantum networks.

In addition, due to the network resource limitation, we also evaluate the satisfying ability of the network to the number of input demands. Moreover, a vital factor affecting the fidelity of the entanglement state between two qubits is also assessed in this section; it is the path length measured by the number of hops from the source to the destination of demands.

\subsection{Performance Metric}
The objective of the problem is to maximize the entangled routing rate, hence a significant metric. In addition, to illustrate the responsiveness of the network, we also employ the number of demands that the network could satisfy simultaneously as an additional metric.

\subsection{Simulation Results}
\subsubsection{Impact of the Number of Input Demands}
\begin{figure}[htbp]
\center
\subfigure[]{\includegraphics[width=4cm]{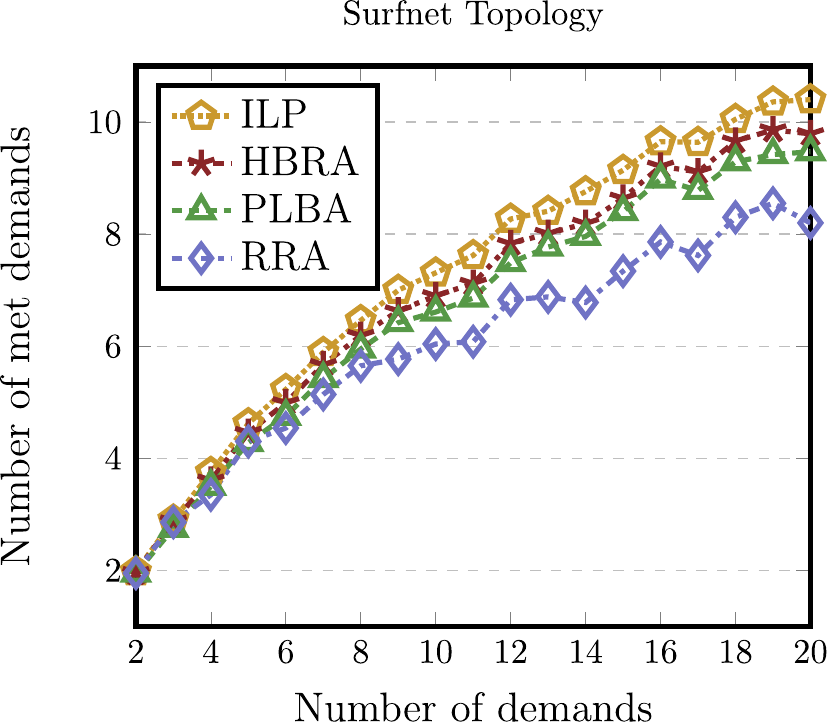}\label{fig:surf_num}}
\subfigure[]{\includegraphics[width=4cm]{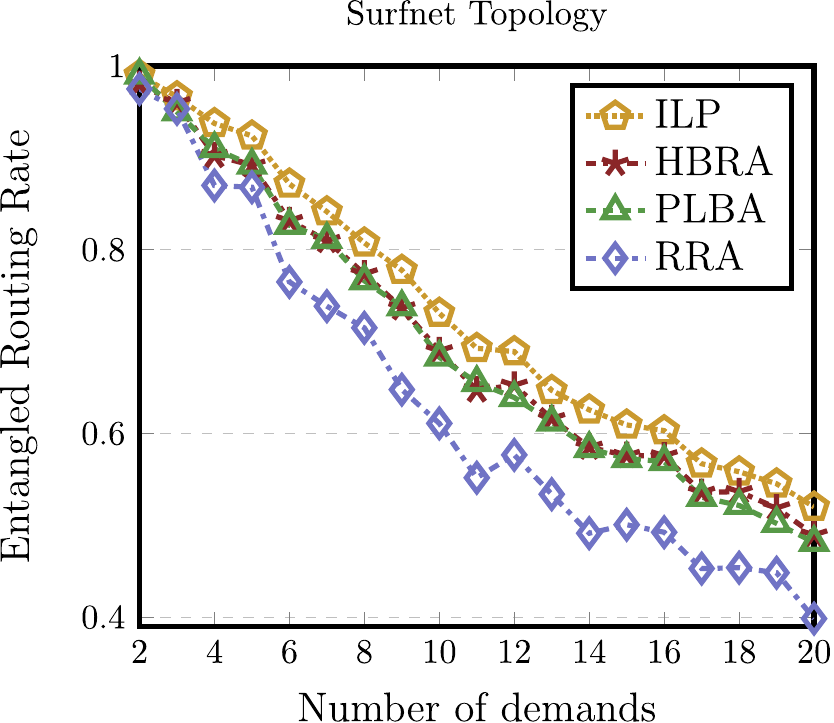}\label{fig:surf}}
\caption{Dependence of (a) the number of met demands and (b) the entangled routing rate on the number of input demands with respect to the Surfnet topology.}
\label{fig:eva_alg}
\end{figure}

\begin{figure}[htbp]
\center
\subfigure[]{\includegraphics[width=4cm]{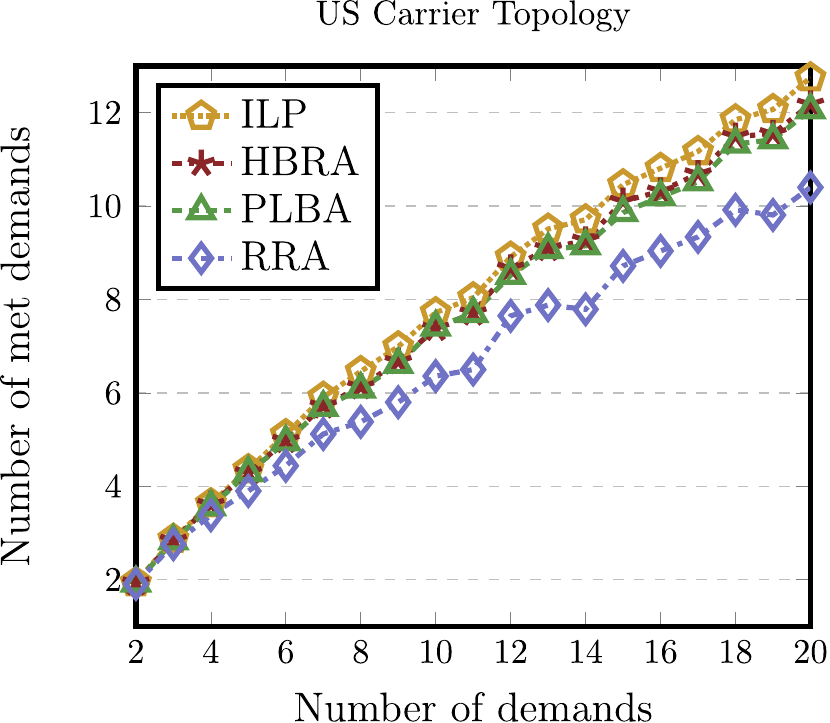}\label{fig:us_num}}
\subfigure[]{\includegraphics[width=4cm]{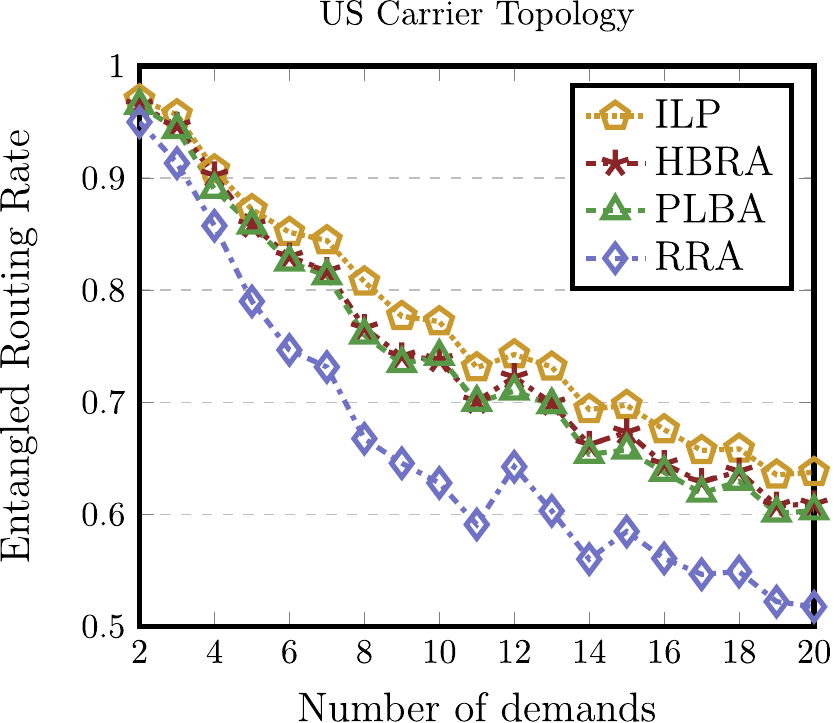}\label{fig:us}}
\caption{Dependence of (a) the number of met demands and (b) the entangled routing rate on the number of input demands with respect to the US Carrier topology.}
\label{fig:eva1_alg}
\end{figure}

To evaluate the effect of the number of input demands, we carry out the simulation on two topologies as mentioned above. The path length of demands is bounded by $8$, which means the length of the met demands does not exceed $8$ hops. From Fig. \ref{fig:surf} and \ref{fig:us}, the entangled routing rate decreases gradually along with increasing the number of input demands, which is equivalent to reducing the satisfying ability of the network in both cases. The reason is that the network resources are exhausted when it has to respond to too many requests at the same time. Between the algorithms, with respect to the final result, the ILP demonstrates the best performance; it provides the optimal solution, which means the maximum ability of the network. The HBRA and PLBA also show their effectiveness; though they are the approximation algorithms, their solutions approach the optimal one with relatively low error. In some situations, the HBRA is a little better than the PLBA. The RRA is a randomized algorithm, hence less stable than the others. It is only effective with a small number of input demands. Figures \ref{fig:surf_num} and \ref{fig:us_num} show more clearly the performance of the algorithms when directly representing the number of met demands by the number of input demands in each topology.

\subsubsection{Impact of the Network Resources}
\begin{figure}[htbp]
\center
\subfigure[]{\includegraphics[width=4cm]{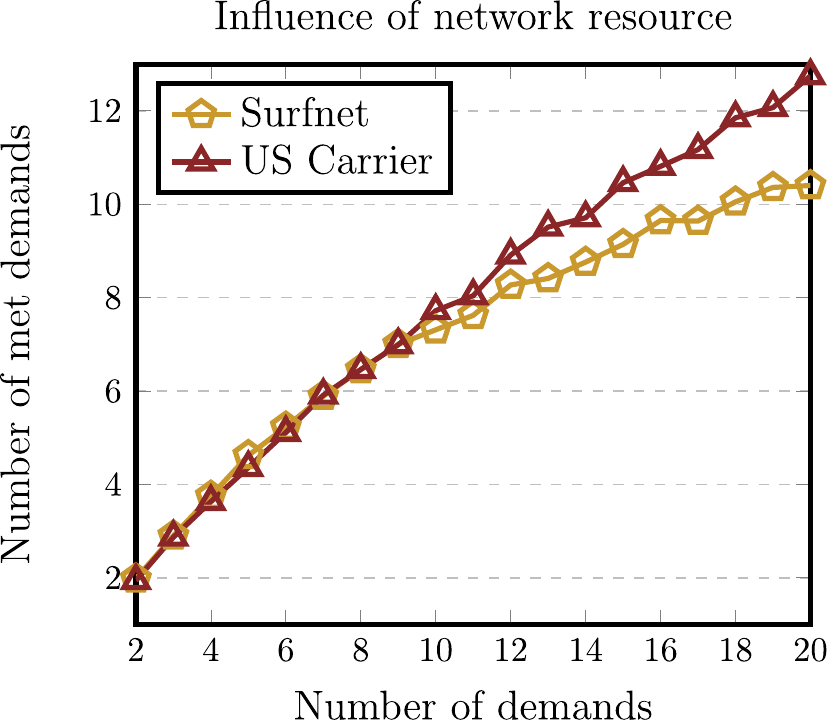}\label{fig:surfnet_vs_us_num}}
\subfigure[]{\includegraphics[width=4cm]{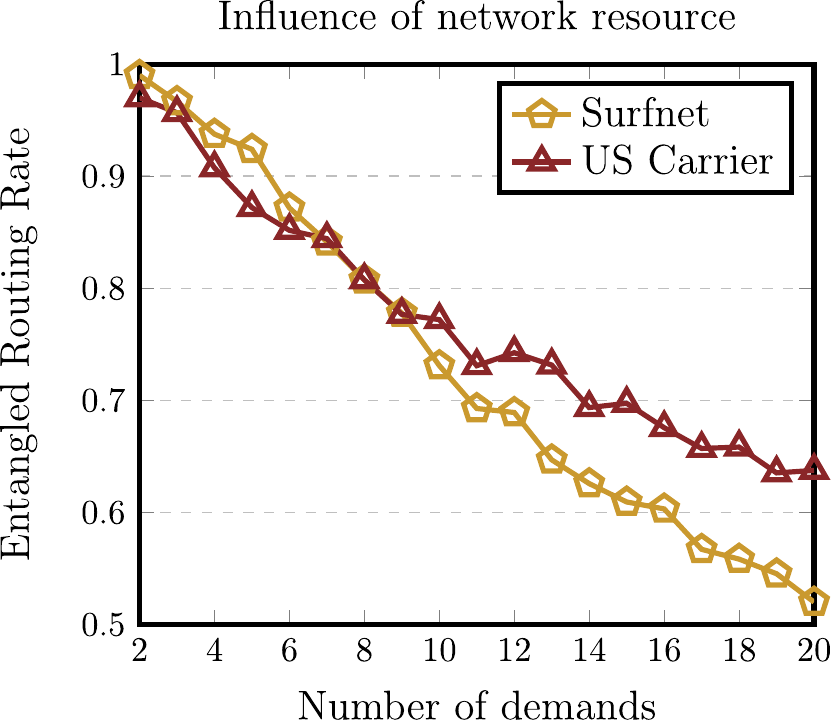}\label{fig:surfnet_vs_us}}
\caption{Dependence of the entangled routing rate on the number of demands (a) Constraint of network resource and (b) Constraint of path length.}
\label{fig:eva_resource}
\end{figure}

In this scenario, we use the ILP to evaluate the influence of the network resources on its satisfying ability. The upper bound of path length is set to $8$. Two networks under examination are Surfnet and US Carrier, as introduced in the above section. With more edges than Surfnet, the responsiveness of US Carrier is better than Surfnet, particularly in the case of increasing the number of demands. As shown in Fig. \ref{fig:surfnet_vs_us}, when increasing the number of demands, the entangled routing rate of both networks declines, but the Surfnet decreases stronger than the US Carrier due to the rapidly exhausting resource, especially when the number of demands is greater than $10$. We can apparently recognize the difference when the number of demands is equal to $20$. The entangled routing rate of US Carrier is about $0.65$, while Surfnet’s is slightly greater than $0.5$. Likewise, Fig. \ref{fig:surfnet_vs_us_num} displays the dependence of the number of met demands on the number of input demands. At the threshold of $10$, while the satisfaction ability of the US Carrier still grows steadily, the corresponding one of the Surfnet also increases but slowly, and it tends to remain unchanged when the number of input demands exceeds $16$ due to depleting the network resources.

\subsubsection{Impact of Path Length}
\begin{figure}[htbp]
\center
\subfigure[]{\includegraphics[width=4cm]{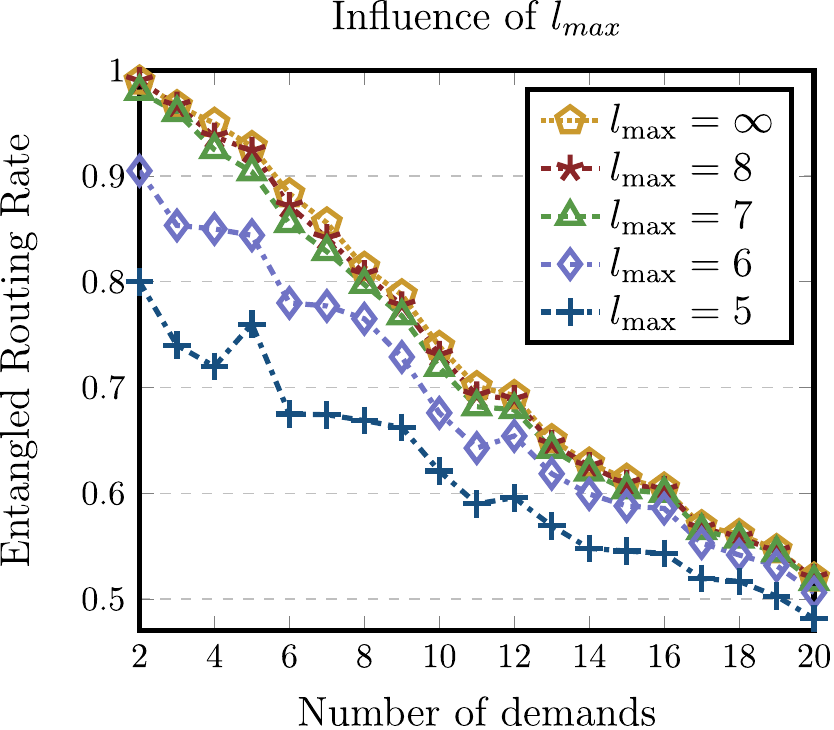}\label{fig:lmax_ratio}}
\subfigure[]{\includegraphics[width=4cm]{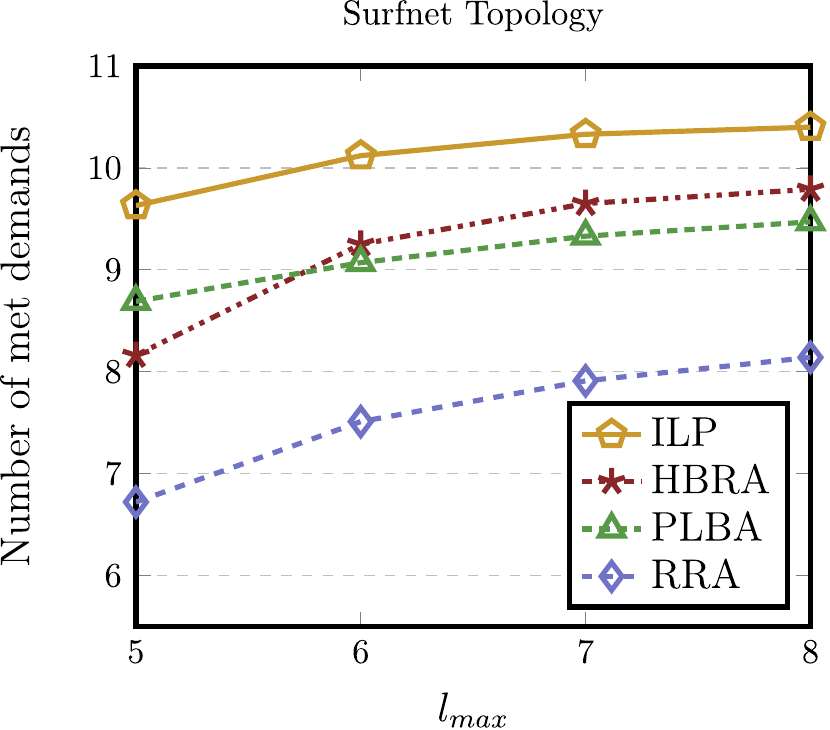}\label{fig:lmax_num}}
\caption{(a) Dependence of the entangled routing rate on the number of demands (b) Dependence of the number of met demands on the upper bound of path length ($l_{\max}$).}
\label{fig:lmax}
\end{figure}

The path length is a significant factor that affects the quality of entanglement. In the quantum network, to guarantee the fidelity of entangled pairs, one has to bound the path length of demands. Nevertheless, this affects the responsiveness of the network. In this part, we assess the impact of the path length on the satisfying ability of the network. In the first scenario, we utilize the ILP to evaluate the influence of the upper bound $l_{\max}$ of the path length on the entangled routing rate of Surfnet. As displayed in Fig. \ref{fig:lmax_ratio}, the constraint of the path length substantially affects the satisfying ability of the network when the upper bound is less than $7$. In particular, when the upper bound is equal to $5$, the entangled routing rate decreases considerably compared with the others. There are two key reasons that can interpret this result. The first one is that the distance in terms of the number of hops between the source and the destination of many demands is too long, or the shortest path between the source and the destination is greater than $5$. This is reasonable in reality because the source and destination are arbitrary nodes in the network. The second reason is that the network resources are not sufficient to serve the demands requiring too strict path length. With an upper bound greater than or equal to $7$, the influence is negligible. Thereby, we see that in order to optimize the number of met demands, the algorithm tends to limit the path length below $8$.

To evaluate the proposed algorithms, we fix the number of input demands to $20$. As shown in Fig. \ref{fig:lmax_num}, all algorithms can improve the number of met demands when increasing the upper bound of path length. Between the proposed algorithms, HBRA provides the best result compared with the optimal solution (ILP). For all algorithms, the number of met demands tends to remain unchanged when $l_{max}$ is greater than or equal to $7$. This is appropriate with the result of the above part.

\section{Discussion: From Theoretical Concepts to Real-world Experiments} \label{sec:discussion}

In the above section, we carried out experiments to evaluate the performance of the proposed methods to solve the MERR problem. We also conducted simulations on the open platform NetSquid to get insight into quantum networks in reality. In this section, we assess the appropriation of the proposed approaches with a real network in two aspects: timing and fidelity. The network we employ in this section is Surfnet. As we stated in the above section, owing to the strict timing of the quantum network, we need methods with extremely fast running time. The requirement is that the interval from beginning the entanglement to the time when the repeaters receive the decision swapping must be less than the decoherence time of qubits, which is about $1.46$ seconds \cite{abobeih2018one} in terms of NV center \cite{childress_hanson_2013}. We can divide the above interval into four smaller ones in order of the progress: (1) the interval for all adjacent nodes performing entanglement, (2) the interval for all nodes in the network sending information about the entanglement states to the centralized controller, (3) the interval for the controller doing routing, (4) the interval for the controller sending the swapping information to all nodes in the network. According to the data we got from the platform in Fig \ref{fig:ev56}, the necessary time for step (1) is about $0.3$ seconds from the first generating of entanglement to the last one in terms of the Surfnet graph. Step (3) is the running time of the algorithms, which is about $0.7170$ seconds for the ILP,  $0.3742$ seconds for the HBRA, and $0.1096$ seconds for the PLBA. We are running for the worst case when there are $20$ demands. The configuration of the computer we employed to perform the simulation is as follows: Intel Core i7-4770, $3.4$GHz, $12$GB memory. Steps (2) and (4) are performed over the traditional network. Currently, the distance between two nodes is not too large, which is about several ten kilometers \cite{10.1145/3465481.3470056}, which corresponds to a communication delay of several hundred nanoseconds, so the required intervals for these steps are negligible compared with steps (1) and (3). In practice, the furthest distance of a quantum channel is now about $4600$km, which connects a station on the ground with a satellite \cite{chen2021integrated}. Nevertheless, we are only concerned with the networks on the ground in this paper. In short, the total required time from step (1) to step (4) never exceeds the threshold ($1.46$ seconds), hence satisfying the requirement of the decoherence time of qubits.

There are many factors that affect the fidelity of end-to-end pairs. As stated in the introduction, we concern the length of the path that connects the source and the destination of a demand. Fig. \ref{fig:ev4} displays the end-to-end fidelity in some circumstances, in which the best is equivalent to the pretty good condition of the quantum channel. That is, both the probability of losing qubit ($5$\%) and the attenuation factor along the channel ($0.025$dB/km) are quite small. With this case, we can achieve the desired threshold of fidelity ($0.5$) when the number of nodes along the path, taking into account the two end nodes, is less than or equal to $8$. Let review Fig. \ref{fig:lmax_ratio}, the number of $8$ nodes is equivalent to $l_{max} = 7$, whose curve nearly approaches the case of $l_{max} = \infty$. That means, even without any constraint of path length, the proposed method is quite good in the aspect of satisfying the lower bound of fidelity. In practice, the condition of quantum channels may be worse than the above \cite{10.1145/3465481.3470056}, which results in lower fidelity. For these situations, we can use the constraint of path length to satisfy the fidelity, but the entangled routing rate will decrease substantially, as shown in Fig. \ref{fig:lmax_ratio}.

\section{Conclusion}\label{sec:conclusion}
In this paper we aim to design approximation algorithms to construct
entangled routing paths connecting all demands in the quantum networks while meeting the network's fidelity constrain. In particular, we have
formulated the problem of maximizing entangled routing rate (MERR) and also demonstrated its hardness as an NP-hard problem.
To obtain the approximation algorithms we have implemented the method using the integer linear programming (ILP) model and rounding techniques.
An ILP employed our designed metrics can optimally decide which qubits to pair and to be included in the routing paths for the small scale network and algorithms using rounding techniques ((HBRA, PLBA) provide good approximation ratios to deal with the challenge of the combinatorial optimization problem in the big scale network.

Using both simulations and experiments with real-world network topologies and traffic matrices, we implemented the proposed algorithms and verified their performances. Both HBRA and PLBA yield pretty good solutions while the ILP can obtain the optimal solution for the small scale network.

\bibliographystyle{IEEEtran}
\bibliography{bib-tn}
\end{document}